\documentclass[article]{revtex4}
\usepackage{epsfig}

\newcommand{\Bo}{\overline{\omega}}
\newcommand{\BO}{\overline{\Omega}}
\newcommand{\Br}{\bar{r}}
\newcommand{\Bs}{\bar{s}}
\newcommand{\Bt}{\bar{t}}
\newcommand{\Ho}{\widehat{\omega}}

\newcommand{\Hr}{{\hat{r}}}

\newcommand{\Ht}{{\skew0{\hat t}{}}}
\newcommand{\To}{\widetilde{\omega}}
\newcommand{\TO}{\widetilde{\Omega}}
\newcommand{\Tlr}{\tilde{r}}
\newcommand{\Ts}{\tilde{s}}
\newcommand{\Tt}{\tilde{t}}

\newcommand{\Sv}{{\mathrm{Sv}}}
\newcommand{\Ss}{{\mathrm{Ss}}}

\newcommand{\ds}{{\mathrm{S3}}}
\newcommand{\fs}{{\mathrm{fs}}}

\newcommand{\RW}{{\mathrm{RW}}}

\newcommand{\beq}{\begin{equation}}
\newcommand{\eeq}{\end{equation}}
\newcommand{\bea}{\begin{eqnarray}}
\newcommand{\eea}{\end{eqnarray}}
\newcommand{\nn}{\nonumber}

\begin{document}

\title{Conformal coordinates for a constant density star}

\author{Karthik Shankar and Bernard F. Whiting}

\address{Department of Physics, PO Box 118440, University of Florida,
Gainesville, FL 32611}

\begin{abstract}
It is well known that the interior of a constant density spherical
star is conformally flat. In this paper we obtain the coordinate system
in which the conformal flatness of the metric manifests itself. In
a similar way, we also construct such coordinates for Robertson Walker
metric.
\end{abstract}
\maketitle

\section{Introduction}

%\footnote{Dedicated to Hans A.\ Buchdahl, whose contributions behind the scene have too often gone unnoticed.}
The constant density, spherically symmetric, perfect fluid solution
%\beq
%\skew0\hat{t}o\skew1\hat{t}o\skew2\hat{t}o\skew3\hat{t}, and \Hst o
%\eeq
to Einstein's equations was given in 1916 by Schwarzschild\cite{key-1}.
Despite being a stalwart of introductory General Relativity, the source
of much of the knowledge we have about the Schwarzschild star interior
solution is shrouded in obscurity. When he was writing about it in
1971, Buchdahl\cite{buchdahl} decried the fact that no standard texts
available to him even mentioned that the metric for this geometry
is conformally flat. To our knowledge, the first documented record
concerning this fact was published only a few years earlier by Shepley
and Taub\cite{shepley}, in a work which Buchdahl actually did not
cite. In fact, both of these papers seem to have been remarkably little
read, or at least referred to, in the, almost forty, intervening %thirty odd 
years.  Like many others from around that time, Shepley and Taub apparently
came across the conformal flatness of the Schwarzsdchild interior
solution in the course of establishing a different, somewhat more
general, geometric result -- including its unique properties %ness 
among perfect
fluid matter sources\cite{buchdahl,raychaudhuri}. For his part, Buchdahl
wrote down coordinates in which the conformal flatness was manifest,
but he neither established their existence specifically nor discussed
their properties in any detail. Instead he concentrated on describing
ensuing optical properties, for the systematic study of which he is
justly held in high regard.

In this paper, we fill in specific %the major 
gaps in Buchdahl's discussion.
We find constructively the flat coordinates in which the conformal
flatness is manifest and we discuss their domain of applicability
and related properties in detail. Forming a backdrop to our presentation
is the need to show a modern application (which will be published
elsewhere) of calculating the self force of a static electric charge
placed inside/outside a Schwarzschild star. Just as importantly, we
also wish to prevent the earlier awareness \cite{buchdahl} of this
conformal flatness and its underlying implications from fading into
oblivion.

The layout of the paper is as follows. We first demonstrate the conformal
flatness of the spatial part of the metric by relating it to the metric
on the 3-sphere. Then we find the family of conformal factors and %two 
related %sets of 
coordinates in which the conformal flatness of the 4-geometry
is manifest, and discuss their domains of applicability.  
%and relate them to the singular limit which arises for the extreme 
%Schwarzschild star, described further below. 
We then briefly address the 
implications of this work for the %particular 
problem of static electromagnetic sources in the interior Schwarzschild star.  
%solution. 
Finally, we mimic our procedure
to indicate the conformal coordinates for another conformally flat metric
(Robertson-Walker).

\section{Preamble}

Consider the metric of a spherically symmetric, constant density star
of mass $M$ and radius $R_{s}$ as given by Schwarzshild\cite{key-1}.
The exterior of the star is just the Schwarzschild metric (where the subscript 
%$\Sv$ 
refers to the {\bf S}chwarzschild {\bf v}acuum solution): 
\begin{equation}
ds^{2}_{\Sv}=-\left(1-\frac{2M}{r}\right)dt^{2}+%\frac{1}
{\left(1-\frac{2M}{r}\right)}^{-1}dr^{2}+r^{2}d\Omega^2.
\end{equation} 
%\left[d\theta^{2}+\sin^{2}\theta\, d\phi^{2}\right].\]
%As we will 
We preserve spherical symmetry throughout, %we 
and so
denote the %metric on the %geometric, unit 
2-sphere by the metric%:
\begin{equation}
d\Omega^{2}=d\theta^{2}+\sin^{2}\theta\, d\phi^{2}.
\end{equation}
The star interior %of the star 
has the following metric (subscript refers to the {\bf S}chwarzschild {\bf s}tar):
\begin{equation}
ds^{2}_{\Ss}=-e^{2\Phi(r)}dt^{2}+%\frac{1}
{\left(1-\frac{2m(r)}{r}\right)}^{-1}dr^{2}+r^{2}d\Omega^2
%\left[d\theta^{2}+\sin^{2}\theta\, d\phi^{2}\right]
,\label{1}
\end{equation}
where $\exp[\Phi(r)]=\frac{3}{2}\sqrt{1-2M/R_{s}}-\frac{1}{2}\sqrt{1-2Mr^{2}/R_{s}^{3}}$
and $m(r)=M\, r^{3}/R_{s}^{3}$.
The quantity $M/R_{s}$ is bounded by $4/9$. For the extreme density
case, when $M/R_{s}=4/9$, the center $r=0$, develops a singularity.

The Weyl tensor for this metric evaluates to zero, %which indicates %
indicating 
that the metric is conformally flat. This implies that, about any
point in the interior of the star, locally (for a finite region) there
exist %s a coordinate system 
coordinates in which the metric would be Minkowski
metric up to a conformal factor, that is $g_{\mu\nu}=\Omega^{2}(x_{\mu})\,\eta_{\mu\nu}$.
In particular, about $r=0$, since we %would also 
expect the spherical
symmetry to be preserved, there should exist coordinates $\{T(t,r),R(t,r),\theta,\phi\}$
such that (\ref{1}) can be rewritten as (subscript $\fs$ refers to {\bf f}lat {\bf s}pace): 
\begin{equation}
ds^{2}_{\Ss}=\Omega^{2}(t,r)\left[-dT^{2}+dR^{2}+R^{2}d\Omega^2
%\left[d\theta^{2}+\sin^{2}\theta\, d\phi^{2}\right]
\right]=\Omega^2(t,r) ds^{2}_{\fs}.\label{conf:eqn}
\end{equation}
This coordinate transformation preserves spherical symmetry (it does
not touch the angular variables). In this paper, we find the coordinates
\{$T(t,r),R(t,r)$\} and the conformal factor $\Omega(t,r)$, which we refer to collectively as the conformal transformation. 
%We shall henceforth term these as the conformal coordinates.

In section \ref{section 1}, we modify the metric (\ref{1}) by %explicitly 
using
a coordinate transformation to express $ds^{2}_{\Ss}$ as $\To^{2}d\Ts^{2}$, where $d\Ts^{2}$ is spatially flat.
In section \ref{section 2}, we work with the metric $d\Ts^{2}$ and find conformally flat coordinates for the extreme density star.
%in the new coordinates (but we call these new coordinates again $t,\, r$).
%In section \ref{section 2}, we 
We also rewrite the metric $d\Ts^{2}$ in a desired
form (presumably used by Buchdahl) which, in Appendix \ref{missed}, we express in terms of a conformal factor $\BO$ and conformal coordinates
\{$T,R$\}, before solving the differential equations they satisfy.
In section \ref{section 3}, we similarly obtain the conformal factor $\BO$, by demanding
that the Riemann tensor of a flat metric vanishes. In section \ref{section 4}, we
use the results from sections \ref{section 2}, \ref{section 3} and appendix \ref{missed} to explicitly construct two sets of conformal coordinates. In section \ref{section 5}, we look into a small application
of the these conformal coordinates in electrostatics. In section \ref{section 6},
we %produce a set of 
examine conformal coordinates for the Robertson-Walker
metric (which is also conformally flat) by following the steps in
section \ref{section 2} through section \ref{section 4}.  Finally, in appendix \ref{appA}, we find the conformal freedom relating (spherically symmetric) flat space to itself (which we use in section \ref{flatcomp}).

%\vfill\eject
\section{Recognizing Buchdahl's contribution}\label{buchdahl}

In his 1971 paper, Buchdahl lists three important results:
\begin{itemize}
\item By using a condition on the Weyl tensor ($C_{klmn}=0$), and by solving an appropriate Einstein equation for the perfect fluid, he established that the Schwarzschild interior solution represented the only static conformally flat distribution of fluid with nonnegative pressure and density.
\item By relating the Schwarzschild star metric to the conformally transformed flat metric, he wrote down the coordinate transformation between the canonical coordinates for the Schwarzschild star and spherical polar coordinates in flat space.  Interestingly for us, he gave virtually no details about his result, but he did include an equation for the flat space orbit of a point on the surface of the star.
\item By introducing (conformally related) coordinates on the 3-sphere, he obtained rather directly an expression for the optical point characteristic --- the time taken by light to propagate between two spatially distinct points.  Throughout, he makes no explicit reference to the fact that the coordinates being introduced reside on the 3-sphere, nor to the fact that spatial slices of the Schwarzschild star are themselves scaled 3-spheres.
\end{itemize}
In this work we extend Buchdahl's analysis in several significant ways:
\begin{itemize}
\item We find the appropriate conformal factor by solving directly $R_{klmn}=0$ for the Riemann tensor of the conformally transformed Schwarzschild interior solution, rather than using the Weyl tensor to find the stellar properties. (section \ref{section 3})
\item We explicitly exhibit the conformal relation to the 3-sphere, and use it to find the 
coordinate transformation to the flat coordinates. (section \ref{section 2} and Appendix \ref{missed})
\item We find the orbit of an arbitrary interior point, and use it to discuss the electromagnetic problem of a static point charge inside the Schwarzschild star, rather than the problem of light propagation. (section \ref{section 5})
\item We fully characterize the parameter freedom in the choice of the conformal factor (and the different domains which arise from it), %especially the multiple domains which arise by different choices of the parameters, 
and demonstrate its $1\!-\!1$ relationship with the conformal freedom to transform flat space into itself. (section \ref{section 4})  
\item To do this, we have characterized the flat space conformal freedom explicitly. (Appendix \ref{appA})
\item We demonstrate the related properties of the Robertson-Walker metrics. (section \ref{section 6})
\end{itemize}
%\vfill\eject

\section{Conformal preparation}\label{section 1}

We first rewrite the interior metric (\ref{1}) as
\beq
ds^{2}_{\Ss}=-\left[a-\frac{1}{2}\sqrt{1-\alpha r^{2}}\right]^{2}dt^{2}+\frac{1}{\left(1-\alpha r^{2}\right)}dr^{2}+r^{2}d\Omega^2
%\left(d\theta^{2}+\sin^{2}\theta d\phi^{2}\right)
,\label{schstr}
\eeq
where $a\equiv\frac{3}{2}\sqrt{1-2M/R_{s}}$ and $\alpha\equiv2M/R_{s}^{3}$:
$a$ can take values in the range $(\frac{1}{2},\frac{3}{2})$. For
the extreme density case, $a=1/2$. Next, we relabel the coordinates
$\sqrt{\alpha}r\rightarrow\Hr$ and $\sqrt{\alpha}t\rightarrow\Ht$.
These new coordinates are dimensionless:
\begin{equation}
\alpha ds^{2}_{\Ss}=-\left[a-\frac{1}{2}\sqrt{1-\Hr^{2}}\right]^{2}d\Ht^{2}+\frac{1}{\left(1-\Hr^{2}\right)}d\Hr^{2}+\Hr^{2}d\Omega^2
%\left(d\theta^{2}+\sin^{2}\theta d\phi^{2}\right)
.\label{2}
\end{equation}
%Observe that, t
The spatial 3-metric corresponds to a 3-sphere, as
can be recognized by writing $\Hr=\sin\eta$
\[
ds^{2}_{\ds}=\frac{1}{1-\Hr^{2}}d\Hr^{2}+\Hr^{2}d\Omega^2
%d\theta^{2}+r^{2}\sin^{2}\theta d\phi^{2}\rightarrow 
\equiv d\eta^{2}+\sin^{2}\!\eta~d\Omega^2
%\left(d\theta^{2}+\sin^{2}\theta d\phi^{2}\right)
.\]
Since the 3-sphere itself is conformally flat, we transform
the coordinates to express this:
\[
ds^{2}_{\ds}=\frac{d\Hr^{2}}{1-\Hr^{2}}+\Hr^{2}d\Omega^2
%\left(d\theta^{2}+\sin^{2}\theta d\phi^{2}\right)
\rightarrow \Ho^{2}(\Hr)\left[d\gamma^{2}+\gamma^{2}d\Omega^2
%\left(d\theta^{2}+\sin^{2}\theta d\phi^{2}\right)
\right].
\]
Because of the spherical symmetry, the above coordinate transformation
will not disturb the angular coordinates $\{\theta,\phi\}$. 
%The conformal factor $\Ho$ depends only on $\Hr$, as does 
The new radial coordinate $\gamma$ depends only on $\Hr$, as does the conformal factor $\Ho$.  %depends only on $r$. 
Equating coefficients of the differentials
in the two forms of the metric gives $d\Hr/\sqrt{1-\Hr^{2}}=\Ho d\gamma$
and $\Hr=\Ho\gamma$, which we solve by eliminating $\Ho$:
\begin{equation}
\frac{d\gamma}{d\Hr}=\frac{\gamma}{\Hr\sqrt{1-\Hr^{2}}}\quad\Rightarrow\quad\gamma(\Hr)=\frac{2\Hr}{1+\sqrt{1-\Hr^{2}}},\hbox{\rm~and so}\label{coord}
\end{equation}
\[
%\Rightarrow 
\Ho(\Hr)=\frac{1+\sqrt{1-\Hr^{2}}}{2},\,\mathrm{~or}\qquad \Ho(\gamma)=\frac{4}{4+\gamma^{2}}.
\]
The interior metric in %of the star 
(\ref{2}) is conformal to a {s}patially {f}lat metric (as indicated in \cite{raychaudhuri}): %(labelled $\FS$ below):
\bea%\[
\alpha ds^{2}_{\Ss}&=&-\left[a-\frac{1}{2}\sqrt{1-\Hr^{2}}\right]^{2}d\Ht^{2}+\Ho^2(\gamma)\left[d\gamma^{2}+\gamma^{2}d\Omega^2
%\left(d\theta^{2}+\sin^{2}\theta d\phi^{2}\right)
\right],\nn\\
%\]
%\[
%\alpha ds^{2}_{\Ss}
&=&-\left[a-\frac{1}{2}\left(\frac{4-\gamma^{2}}{4+\gamma^{2}}\right)\right]^{2}d\Ht^{2}+\frac{16}{(4+\gamma^{2})^{2}}\left[d\gamma^{2}+\gamma^{2}d\Omega^2
%\left(d\theta^{2}+\sin^{2}\theta d\phi^{2}\right)
\right],{\rm~and~so:}\nn
\eea%\]
\[
\frac{(4+\gamma^{2})^{2}}{16}\alpha ds^{2}_{\Ss}=-\frac{1}{64}\left[4(2a-1)+(2a+1)\gamma^{2}\right]^{2}d\Ht^{2}+\left[d\gamma^{2}+\gamma^{2}d\Omega^2
%\left(d\theta^{2}+\sin^{2}\theta d\phi^{2}\right)
\right].
\]
%We again relabel coordinates, $\gamma\rightarrow r$ and $(2a+1)t/8\rightarrow t$, and define $\beta^{2}\equiv4(2a-1)/(2a+1)$, which %we see that it's values
%ranges between $(0,2)$. For the extreme density case, $\beta=0$.
%A conformal factor can be removed to make the metric take the
%following simple form:
We now define $\beta^{2}\equiv(2a-1)/(2a+1)$, which ranges between $(0,1)$, 
being zero for the extreme density case.  %and 
We again relabel coordinates, with $\gamma/2\rightarrow\Tlr$ %$\Tlr$ as given in (\ref{coord})
%$2\Hr/(1+\sqrt{1-\Hr^{2}})\rightarrow \Tlr$ 
and $(2a+1)\Ht/4\rightarrow \Tt$. %For the extreme density case, $\beta=0$.
A conformal factor can be removed to make the metric take the
following simple form:
\begin{equation}
ds^{2}_{\Ss}=\frac{4}{\alpha(1+\Tlr^{2})^{2}}\left[-(\beta^{2}+\Tlr^{2})^{2}d\Tt^{2}+d\Tlr^{2}+\Tlr^{2}d\Omega^2
%\left(d\theta^{2}+\sin^{2}\theta d\phi^{2}\right)
\right]=\To^{2}d\Ts^{2},\label{3}
\end{equation}
where $\To^{2}=4/\alpha(1+\Tlr^{2})^{2}$. Note that these new coordinates
are dimensionless and the conformal factor in front of $d\Ts^{2}$
has the dimensions of $\alpha^{-1}$, which is $\mathrm{[L]}^{2}$.

In this coordinate system, let the radius of the star be denoted as
$r_{s}$. The coordinate $\Tlr$ in metric (\ref{3}) ranges from $0$
to $r_{s}$. Tracing back the coordinate transformations performed
so far, we can express $r_{s}$ in terms of $R_{s}$ as:
\begin{equation}
r_{s}=\frac{\sqrt{\frac{2M}{R_{s}}}}{1+\sqrt{1-\frac{2M}{R_{s}}}}.\label{coord1}
\end{equation}
No matter how big $R_{s}$ is, $r_{s}$ is always less than $1/\sqrt{2}$.
For the extreme density star, $r_{s}=1/\sqrt{2}$. We can express $r_{s}$
and $\beta$ purely in terms of each other as given below:
\begin{equation}
\beta^{2}=\frac{(1-2r_{s}^{2})}{2-r_{s}^{2}},{\rm~and}\qquad r_{s}^{2}=\frac{4(1-2\beta^{2})}{2-\beta^{2}}.\label{coord2}
\end{equation}

\section{Conformal solution}\label{section 2}

The metric $d\Ts^{2}$ given in (\ref{3}) can be expressed as manifestly
conformally flat, with the help of a spherically symmetric conformal
factor $\TO$ and new conformal coordinates $\{T,R\}$: 
%\begin{equation}
\begin{eqnarray}%{cc}
d\Ts^{2}&=&-[\beta^{2}+\Tlr^{2}]^{2}d\Tt^{2}+d\Tlr^{2}+\Tlr^{2}d\Omega^2\nonumber\\
%\left(d\theta^{2}+\sin^{2}\theta d\phi^{2}\right)\\
&=& \widetilde\Omega^{2}(\Tt,\Tlr)\left[-dT^{2}+dR^{2}+R^{2}d\Omega^2
%\left(d\theta^{2}+\sin^{2}\theta d\phi^{2}\right)
\right].\label{4}
\end{eqnarray}%\end{equation}
Consider first the extreme density case, when $\beta=0$. It
is straightforward to obtain: 
\begin{equation}
T(\Tt,\Tlr)=\Tt,\qquad R(\Tt,\Tlr)=1/\Tlr,\qquad\Omega(\Tt,\Tlr)=\Tlr^{2}.\label{35}
\end{equation}
The $(\Tlr,\Tlr)$ component of the Einstein tensor blows up at $\Tlr=0$,
exhibiting a singularity. Hence, performing a coordinate transformation
at $\Tlr\!\!=\!\!0$ is really pointless. We constrain our coordinate
transformation so as to exclude $\Tlr=0$. We let %us say that
the coordinate transformation (\ref{35}) be valid for $\Tlr>r_{0}$,
for some small $r_{0}>0$, and for all values of $\Tt$.

For $\beta\neq0$, there is a further simplification which can be easily made.  
The metric
\begin{equation}
d\Ts^2=-(\beta^2+\Tlr^2)^2 d\Tt^2+d\Tlr^2+\Tlr^2 d\Omega^2,\label{35a}
\end{equation}
is again conformal to a 4-metric with 3-spheres as spatial slices: %gives:
\begin{equation}
d\Ts^2=\Bo^2 %{(\beta^2+\Tlr^2)^2\over 4\beta^2}\times
\Big[-d\Bt^2+d\Br^2+\sin^2(\Br) d\Omega^2\Big]=\Bo^2 d\Bs^2,{\rm~where}\label{3-sphere}
\end{equation}
%where
\begin{eqnarray}
\Bt&=&2\beta\Tt,\qquad\Bo=(\beta^2+\Tlr^2)/2\beta,\quad{\rm and}\label{eqn15}\\
\Br&=&2\arctan(\Tlr/\beta)\le{\cal R}_{s},{\rm~and}\qquad{\cal R}_{s}=2\arctan(r_{s}/\beta).\label{eqn16}
\end{eqnarray}
Then, as shown in appendix \ref{missed}, and where we chose the $\mp$ to preserve time orientation:
\begin{eqnarray}
d\Bs^2&=&\BO^2\left[-dT^{2}+dR^{2}+R^{2}d\Omega^2\right], %ds^2_{\fs}
{\rm~where}\label{17m}\\
\BO&=&\bar{c}\left[\cos(\Br)-\cos(\Bt)\right],\label{17o}\\
%{\rm~where}\qquad d\Bs^2=\BO^2 ds^2_{\fs}\label{17om}\\
R&=&\sin(\Br)/\BO,%\quad
{\rm~and}\label{17r}\\
T&=&\mp\sin(\Bt)/\BO,\label{17t}
\end{eqnarray}
exemplify a transformation to flat space  (with $t_0\!=\!0$), where $\bar{c}$ is selected so that $\BO>0$.  %It is so simple.

%are no such obvious conformal coordinates
%as eq(\ref{35}). After a detailed analysis, we shall show in section
%4 that the conformal factor and coordinates are given by eq(\ref{20}\ref{21})
%and eq(\ref{25},\ref{26}).

%Observe that, $T$ does not appear explicitly in these equations.
%We can algebraically solve for $T'$, $\dot{T}$ and $\Omega$ from
%the last three equations and plug them into the first equation to
%get a partial differential equation for $R(t,r)$ alone.

%\begin{equation}
%\dot{R}^{2}=(\beta^{2}+r^{2})\left[R'^{2}-\frac{R^{2}}{r^{2}}\right].\label{7}\end{equation}

%
%If we have a solution to this differential equation, then we can use
%(\ref{6} I, II) to get solutions for $T(t,r)$ the following way.

%\begin{equation}
%T'^{2}=R'^{2}-\frac{R^{2}}{r^{2}}\,\,\Rightarrow T(t,r)=\int\sqrt{R'^{2}-\frac{R^{2}}{r^{2}}}dr\:+g_{1}(t),\,\,\,\,\mathrm{or}\label{8}\end{equation}
%\begin{equation}
%\dot{T}^{2}=\dot{R}^{2}+\frac{R^{2}}{r^{2}}(\beta^{2}+r^{2})^{2}\,\,\Rightarrow T(t,r)=\int\sqrt{\dot{R}^{2}+\frac{R^{2}}{r^{2}}(\beta^{2}+r^{2})^{2}}dt\:+g_{2}(r).\label{9}\end{equation}
%An alternative to directly solving eq(\ref{7}) is to first find $\Omega(t,r)$
%(as in section 3), and then use (\ref{6} III) to find $R(t,r)$.

\section{Curvature equations}\label{section 3}

From (\ref{17m}), it is clear that all the components of the Riemann
tensor of the following metric should vanish:
\[
%\frac{1}
ds^2_{\fs}={\BO(\Bt,\Br)}^{-2}\left[-d\Bt^{2}+d\Br^{2}+\sin^{2}\!{(\Br)}\,d\Omega^{2}\right].
\]
%Define $a(\Br)\equiv\beta^{2}+r^{2}$. 
In this section, we shall solve for the conformal factor $\BO(\Bt,\Br)$
for which the Riemann tensor vanishes. Evaluating the Riemann tensor
reveals that there are 5 independent non vanishing components $R_{\Bt\Br\Bt\Br}$, $R_{\Bt\theta\Bt\theta}=\sin^{2}\theta$$R_{\Bt\phi\Bt\phi}$, $R_{\Bt\theta\Br\theta}=\sin^{2}\theta R_{\Bt\phi\Br\phi}$,
$R_{\Br\theta\Br\theta}=R_{\Br\phi\Br\phi}$, $R_{\theta\phi\theta\phi}$.
Equating each of these components to zero gives us 5 equations (in this section only, dot denotes a $\Bt$-derivative and prime an $\Br$-derivative):
\[
\begin{array}{cc}
\BO^{4}R_{\Bt\Br\Bt\Br}=\left[\BO'^{2}-\BO\,\BO''-\dot{\BO}^{2}+\BO\,\ddot{\BO}\right]=0, & (\mathrm{i})\\
\BO^{4}R_{\Bt\theta\Bt\theta}=\sin^2(\Br)\left[\BO\,\ddot{\BO}+\BO'^{2}-\BO\,\BO'\cot(\Br)-\dot{\BO}^{2}\right]=0, & (\mathrm{ii})\\
\BO^{4}R_{\Bt\theta\Br\theta}=\sin^{2}(\Br)\BO\,\dot{\BO}'=0, & (\mathrm{iii})\\
\BO^{4}R_{\Br\theta\Br\theta}=\sin^2(\Br)\left[-\BO'^{2}+\BO\,\BO'\cot(\Br)+\BO\, \BO''+\dot{\BO}^{2}+\BO^{2}\right]=0, & (\mathrm{iv})\\
\BO^{4}R_{\theta\phi\theta\phi}=\sin^{4}(\Br)\sin^{2}\theta\left[\dot{\BO}^{2}-\BO'^{2}+2\BO\,\BO'\cot(\Br)+\BO^2\right]=0. & (\mathrm{v})
\end{array}
\]
We now solve these five equations for $\BO(\Bt,\Br)$ in a sequence of simple steps.

\begin{itemize}
\item We first note that these five equations are not all algebraically
independent.  %of each other. 
Elimination of $\dot{\BO}^{2}$ from (iv)
and (v) gives 
{\protect
\[
\begin{array}{cc}
\BO'\cot(\Br)-\BO''=0, & (\mathrm{vi})
\end{array}
\]}
which also follows by eliminating $\ddot{\BO}$ from (ii) and (i). 
\item It is similarly useful to simplify (i) and (iv), or equivalently (ii) and (v) to give: 
\[
\begin{array}{cc}
\ddot{\BO}+\BO'\cot(\Br)+\BO=0. & (\mathrm{vii})
\end{array}
\]
% Subtracting (vi) from (vii), gives an equation for $a$ alone, $a''r-a'=0$,
%which is the requirement for the conformal flatness of the metric
%(\ref{4}). Since $a(\Br)=\beta^{2}+r^{2}$, this equation is automatically
%satisfied. Hence eq(vi) and eq(vii) are identical. 
Thus, to solve for $\BO(\Bt,\Br)$,
it is sufficient to use just the %four independent 
equations (i),(iii),(vi),(vii).
\item The simplest of these equations is (iii). Integrating
(iii) gives $\BO$ in terms of two arbitrary functions, $F(\Bt)$
and $G(\Br)$:
\begin{equation}
\BO(\Bt,\Br)=F(\Bt)+G(\Br).\label{10}
\end{equation}
\item Substitution of this result into (vi) and then integrating gives:
\begin{equation}
G(\Br)=A+B\cos(\Br).\label{11}
\end{equation}
% Comparing the solutions from (\ref{10}) and (\ref{11}), we can
%conclude that $G(\Br)$ is a constant. This leaves us with 
\item Substitution of these results into (vii), and integrating, gives:
\begin{equation}
F(\Bt)=C\cos(\Bt-t_0)-A.\label{12}
\end{equation}
\item Finally, substitution of all the results (\ref{10}) through (\ref{12}) into (i) determines 
%We have so far used only two of the four independent equations. Using
%(\ref{12}), we shall now rewrite (i) and (v) to solve for $F(\Bt)$
%and $C$. Eq(v) takes the form 
\begin{equation}
C=\pm B.\label{13}
\end{equation}
\end{itemize}
%so that, 
So, in the end we have, simply:
\begin{equation}
\BO(\Bt,\Br)=B[\cos(\Br)\pm\cos(\Bt-t_0)].\label{14}
\end{equation}
The form of the time dependence, $(\Bt-t_0)$, signifies the time translation invariance of the metric (\ref{17m}).  In principle, the $\pm$ sign could be eliminated since a shift of $t_0$ by $\pm\pi$ would change the sign of the second cosine term.  Keeping the $\pm$ sign means $t_0$ effectively lies in $[0,\pi)$.  The scaling of $B$ is really immaterial to the conformal transformation, though it is useful to retain, while the sign of $B$ must match the sign of the expression inside the square brackets in (\ref{14}), to ensure that $\BO(\Bt,\Br)$ is always positive.  This means that the family of conformal factors parameterized by $B$ and $t_0$ in (\ref{14}) %domain of $\BO(\Bt,\Br)$ 
really splits into two subfamilies, depending on the sign of the square bracket in (\ref{14}) or, equivalently, the sign of $B$.  For each $t_0$, the members of each subfamily are bounded by (null) surfaces upon which $\BO(\Bt,\Br)=0$.  These separate the domains in which $B$ is positive from those in which $B$ is negative.  %For each $t_0$, these 
Taken together, the different domains from the two subfamilies 
are complementary, in %the sense 
that %, together, 
they then
cover the entire spacetime.

\subsection{Flat space comparison}\label{flatcomp}

In appendix \ref{appA}, %the appendix, 
we study the properties of the most general spherically
symmetric conformal transformations on flat space which can be obtained
by a coordinate transformation. The results are summarized in (\ref{A2}).
We see there that the conformal factor $H(T,R)$ also has two free parameters,
$D$ and $T_{0}$.  
A unique relationship exists between
these two parameters and the changes they induce in the two parameters $\{B,t_{0}\}$ in $\BO(\Bt,\Br)$ as given in 
(\ref{14}). We now find this relationship.  %in the following way.

Consider the metric $d\Ts^{2}$ in (\ref{4}). Let us transform
the conformal coordinates ($T,R$) to new set of coordinates ($T',R'$)
such that a conformal factor $H(T,R)$ is pulled out as in shown in (\ref{A1}): 
\begin{equation}
d\Ts^{2}=\BO^{2}\left[-dT^{2}+dR^{2}+..\right]=\BO^{2}H^{2}\left[-dT'^{2}+dR'^{2}+..\right]\label{e1}
\end{equation}
Since (\ref{14}) gives the most general form of $\BO$, we know
%can be sure 
that, for any given $\BO(\Bt,\Br)$ (fixed $B,t_{0}$) and
$H(T,R)$ (fixed $D,T_{0}$), the factor $\BO H$ should be of
the same form as the $\BO$ given in (\ref{14}), %and 
in which %the 
corresponding parameters 
$B'$ and $t_{0}'$ can be uniquely determined in terms of $\{B,t_{0}\}$ and
$\{D,T_{0}\}$. 
%Without loss of generality, we can choose $\BO(\Bt,\Br)$ with $A=1$ and $t_{0}=0$ and we can also choose $G(T,R)=R^{2}-(T-T_{0})^{2}$.
%{\it Corresponding to $\BO(\Bt,\Br)$, we can compute the conformal coordinates
%$T(\Bt,\Br)$ and $R(\Bt,\Br)$ by following the method shown in section \ref{section 4}.
%This would let us compute $H$ as a function of $t,r$.} %Further c
%Calculation show that t
Thus, the product $\BO H$ must take the form,
\begin{equation}
\BO H=B'\left[\cos(\Br)\pm\cos(\Bt-t_{0}')\right]\equiv\BO'.%{\rm~where}
\label{e2}
\end{equation}
It can be shown that %where
\begin{equation}
B'=\frac{(B^{2}T_{0}^{2}+1)D}{B},{\rm~and}\qquad t_{0}'=t_0\pm2\arctan\left({1\over BT_{0}}\right).
\label{e3}
\end{equation}
Note that the parameter $T_0'$ of (\ref{A2}) has also been changed, as similarly occurs in (\ref{A4}):
\begin{equation}
\delta T_0'= \frac{B^2T_0}{(B^2T_0^2+1)D}.\label{e4}
\end{equation}
Inverting, we can find $\{D,T_0\}$, for a given $\{B,t_0\}$, in terms of any chosen $\{B',t_0'\}$:
\begin{equation}
D=BB' \sin^2[(t_0'-t_0)/2],{\rm~and}\qquad T_0=\pm{\cot[(t_0'-t_0)/2]\over B}.
\end{equation}
%Observe that, by a suitable redefinition of the $A'$ and $t_{0}'$,
%the product $\BO G$ (eq\ref{e2}) can be reformed to appear as
%$\BO$ in eq(\ref{19}). We have specifically chosen $A=1$, $t_{0}=0$
%and $C=1$. Any other choice of these constants would make trivial
%changes in our final result (eq\ref{e3}) like scaling $A'$ by a
%factor $AC$ and adding a constant $t_{0}$ to $t_{0}'$. Thus, we deduce that 
Clearly, every member in the family of conformal factors $\BO$ 
given in (\ref{14})
can be obtained from just one of their representatives through %flat space 
conformal
%by performing coordinate 
transformations as shown in appendix \ref{appA}.  %the appendix in (\ref{A1}).
%In other words, t
The degrees of freedom %that are seems to be 
available for choosing a conformal factor of the form (\ref{14}) entirely
correspond to the degrees of freedom available in choosing 
a conformal factor $H$ which still maintains 
a flat coordinate system, as shown in appendix \ref{appA}.  %the appendix.

\section{Stellar coordinate domains}\label{section 4}

In this section, we consider two specific forms of $\BO(\Bt,\Br)$
(one from each subfamily), by suitably choosing the $\pm$ sign, and the parameters $B$ and $t_{0}$ in (\ref{14}). We then construct the conformal
coordinates $R(\Bt,\Br)$ and $T(\Bt,\Br)$ %(see section \ref{section 2}) 
corresponding to each of these $\BO$. 
%We observe that, when $\beta\ne 0$ %is nonzero, 
One of these coordinate transformations (part A) is well defined everywhere
within a given range of $\Bt$ around $\Bt=t_0$, while the other set of
coordinates (part B) is never well defined at the origin $\Br=0$.
%We will also observe that in the limit $\beta\rightarrow0$, 
For the coordinates in part B, the limit $\beta\rightarrow0$ will 
correspond to (\ref{35}).  %, while the other 
The set of coordinates in part 
A does not have such a well defined limit. In part C, we provide
the inverse coordinate transformations corresponding to the coordinate
transformations of part A.

\subsection*{A.}\label{section 4A}

First, we consider $\BO$ as given in (\ref{14}) with $B=1$, the $+$ve sign taken, and $t_{0}=0$: 
\beq
\BO(\Bt,\Br)=\cos(\Br)+\cos(\Bt).\label{20}
\eeq
From (\ref{6} III), we obtain $R(\Bt,\Br)$ %=\sin(\Br)/\BO(\Bt,\Br)$, 
while from (\ref{17t}) or (\ref{missed15}) we take $T(\Bt,\Br)$ and write: 
\beq
R(\Bt,\Br)={\sin(\Br)\over\BO(\Bt,\Br)},{\rm~and}\qquad T(\Bt,\Br)={\sin(\Bt)\over\BO}.\label{23}  
\eeq
%. This implies\begin{equation}
%R(\Bt,\Br)=\frac{r}{\beta^{2}-(\beta^{2}+r^{2})\,\sin^{2}(\beta t)}.\label{20}\end{equation}
%%One can easily check that (\ref{20}) is a solution to (\ref{7}).
%We shall now use (\ref{8}) to find the solution for $T(\Bt,\Br)$.
%This yields\begin{equation}
%T(\Bt,\Br)=\frac{\beta\cot(\beta t)}{\left[\beta^{2}-(\beta^{2}+r^{2})\,\sin^{2}(\beta t)\right]}+g_{1}(\Bt).\label{21}\end{equation}
%Using this form of $T(\Bt,\Br)$ and $R(\Bt,\Br)$ in (\ref{6} IV), we
%end up with a simple differential equation for $g_{1}(\Bt)$.
%\begin{equation}
%\frac{dg_{1}}{dt}=\frac{1}{\sin^{2}(\beta t)}\Rightarrow g_{1}(\Bt)=-\frac{\cot(\beta t)}{\beta}-T_{0}.\label{22}\end{equation}
%where $T_{0}$ is an arbitrary constant which signifies the time translation
%invariance of the metric (\ref{4} in the conformal coordinates.
%Let us set $T_{0}=0$. Substituting (\ref{22} in (\ref{21}, we
%obtain \begin{equation}
%T(\Bt,\Br)=\frac{(\beta^{2}+r^{2})\tan(\beta t)}{\beta\left[\beta^{2}-r^{2}\tan^{2}(\beta t)\right]}.\label{23}\end{equation}
%These coordinate transformations are valid only locally. To find out
%the range covered by this coordinate system, we note that the coordinate
%patch is good only as long as the denominator in the coordinate functions
%and the conformal factor are non vanishing. 
For each $\Br$, the coordinate
transformation is locally valid for $\Bt$ in the range (see Fig \ref{fig2}):  %as long as 
\begin{equation}
%-\frac{1}{\beta}\mathrm{tan}^{-1}\left(\frac{\beta}{r}\right)<t<\frac{1}{\beta}\mathrm{tan}^{-1}\left(\frac{\beta}{r}\right)
-\pi+\Br<\Bt<\pi-\Br.\label{24}
\end{equation}
Nevertheless, this transformation maps %the allowed 
region 1 of Fig \ref{fig2} into the whole of flat space.  Since $\Br\leq {\cal R}_{s}$, $\pi-{\cal R}_{s}$ is a lower bound for $\pi-\Br$.
Hence, the range of unconditional validity %of the coordinate transformation 
inside the star is a
strip of finite width in the t-coordinate: $\left|\Bt\right|<\pi-{\cal R}_{s}$.

%In the limit $\beta\rightarrow0$, the width of this strip is $1/r_{s}=1/\sqrt{2}$.
%That is, only for $\left|t\right|<1/\sqrt{2}$, the coordinate transformation
%is valid. But this coordinate transformation (\ref{20}, \ref{23})
%does not reduce to (\ref{35} in the limit $\beta\rightarrow0$.
%In fact, the limit is not even well defined.

\subsection*{B.}

\begin{figure}[htbp] 
\scalebox{0.50}{ \psfig{file=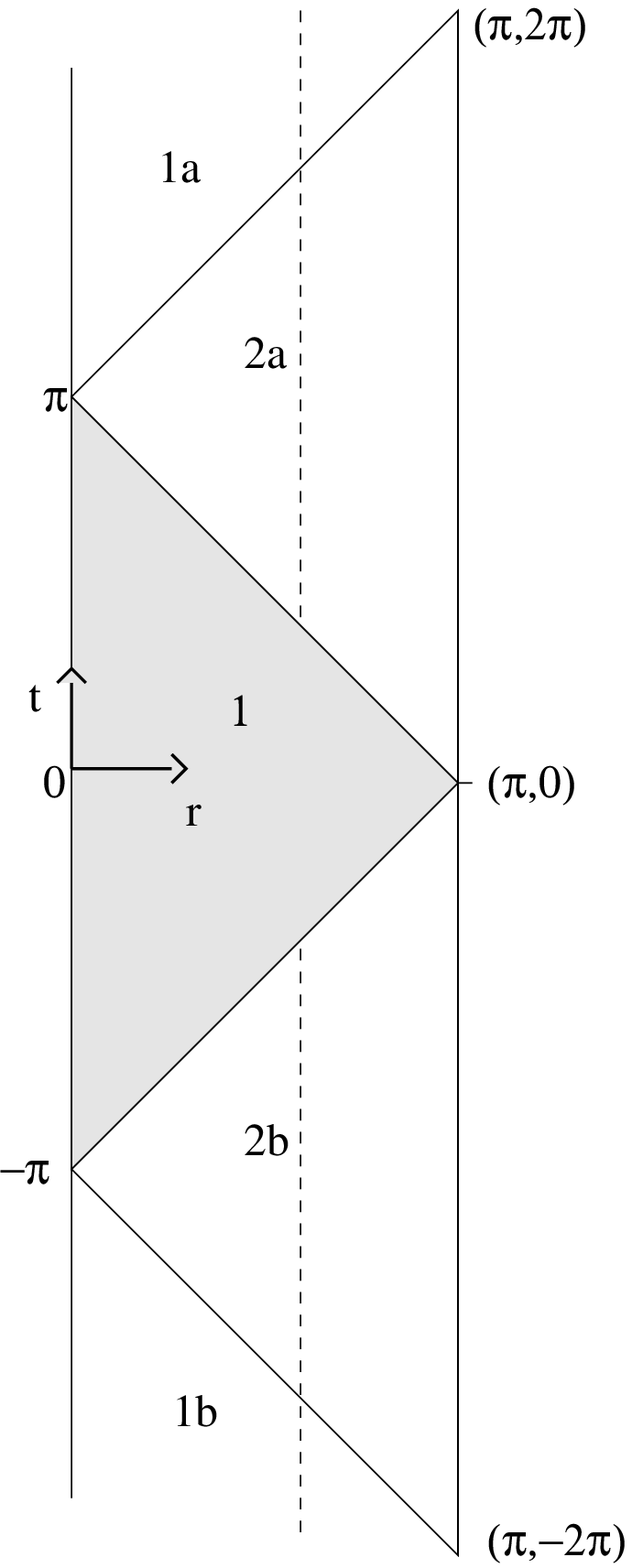} }
\hbox to 2.70truein{}
\scalebox{0.50}{\psfig{file=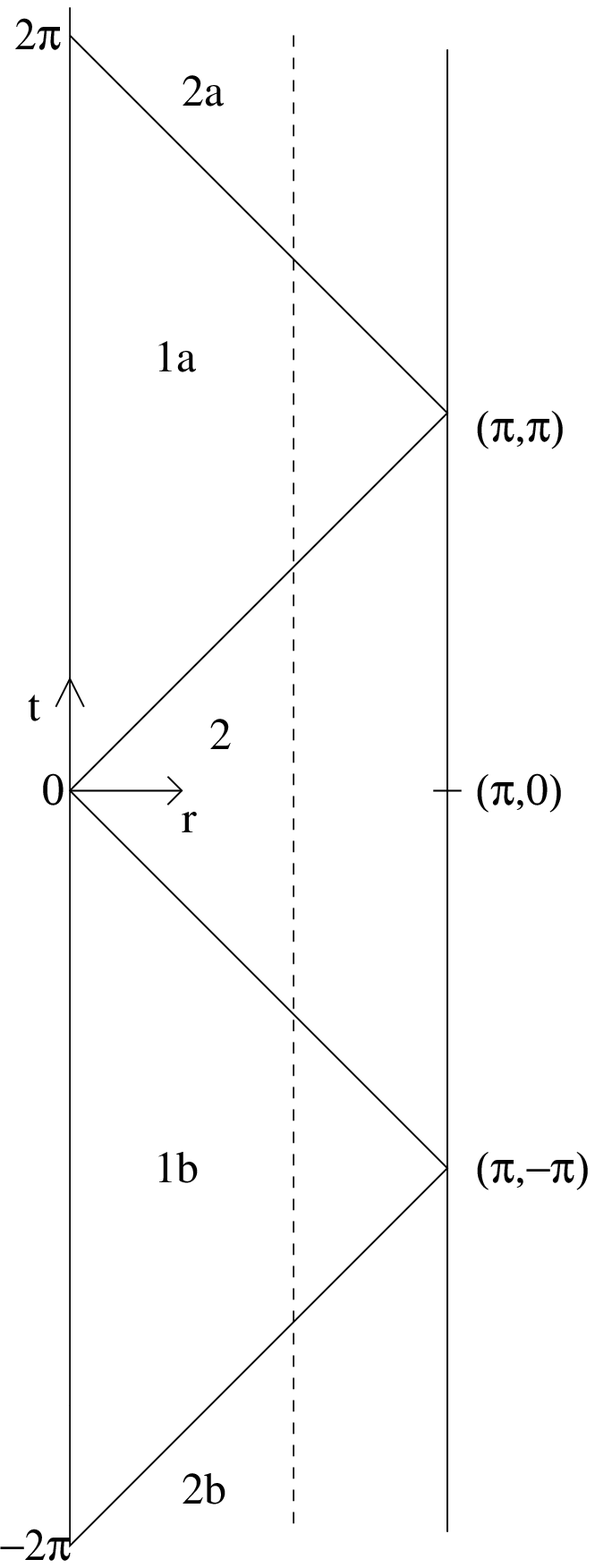} }
\caption{On the left, the region mapped by $\BO$ given %the transformation 
in part A is shown shaded, and labeled 1.  Regions 1a, 1b are also mapped by this transformation.  To map regions 2a, 2b requires an overall sign change in (\ref{20}).  On the right, the regions mapped by $\BO$ in part B are labeled 2, and 2a, 2b.  Regions 1a, 1b require an overall sign change in (\ref{25}).}
\label{fig2} 
\end{figure}

Next, we choose %make a suitable choice of 
parameters $\{\pm,B,t_{0}\}$ in (\ref{14})
to give %express 
the conformal transformation as:  %$\BO(\Bt,\Br)$ as
\begin{equation}
\BO(\Bt,\Br)=\cos(\Bt)-\cos(\Br),{\rm~and}\label{25}
\end{equation}
%We proceed in the same way as before to get a different set of conformal
%coordinates 
\begin{equation}
R(\Bt,\Br)=\frac{\sin(\Br)}{\BO},\qquad T(\Bt,\Br)=\frac{\sin(\Bt)}{\BO}.\label{26}
\end{equation}
For each $\Br$, the coordinate transformation (\ref{26}) is valid for $\Bt$ in the region $-\Br<\Bt<\Br$  (see Fig \ref{fig2}).
%\begin{equation}
%-r<t<r.\label{27}
%\end{equation}
%Nevertheless, t
This transformation similarly %also 
maps %the allowed 
region 2 in Fig \ref{fig2}
into the whole of flat space.  Unlike in the previous case (A), there exists no nonzero lower bound
for $\Br$. 

At this point we restore the $\beta$ dependence by reintroducing (\ref{eqn15},\ref{eqn16}), along with the coordinates of (\ref{35a}).  
%Before taking the limit $\beta\rightarrow0$, we must first substitute $\BO=?$  %Hence, f
For $\beta$ away from zero, the coordinate transformation (\ref{26})
%system 
is not valid at the origin, %in a region that contains 
$\Br=0,\Bt=0$.  
Recall that in section \ref{section 2}, for $\beta=0$, %case (\ref{35}), we assumed
we took the coordinate transformation (\ref{35}) to apply only in the
region $\Tlr>r_{0}$. Similarly, by restricting $\Tlr$ to be greater than
$r_{0}$  when considering (\ref{eqn15}) and (\ref{eqn16}) in the coordinate transformation in (\ref{26}), we have a
nonzero lower bound for $\Tlr$, and thus, in the coordinates of (\ref{35a}): 
%Hence, from (\ref{27}), it is
%clear that the coordinate transformation (\ref{26}) is valid for
\begin{equation}
-\frac{1}{\beta}\mathrm{tan}^{-1}\left(\frac{r_{0}}{\beta}\right)<\Tt<\frac{1}{\beta}\mathrm{tan}^{-1}\left(\frac{r_{0}}{\beta}\right).\label{28}
\end{equation}
In the limit $\beta\rightarrow0$, it can be shown that %(\ref{25}, 
(\ref{26}) reduces to (\ref{35}) while %and 
the region of validity of the coordinate
transformation (\ref{28}) becomes $\left|\Tt\right|<\infty$, 
%which essentially means that the coordinate transformation is valid for
%all values of $t$, 
exactly as in (\ref{35}).

\subsection*{C.}

Since the conformal coordinates given in part A %(\ref{20}, 
(\ref{23}) have $r\!=\!0\!\Leftrightarrow\!R\!=\!0$ for all $\{T,t\}$, and
are well behaved near $r\!=\!0$, it is possible to obtain the inverse
coordinate transformations $r(T,R),t(T,R)$ there. In fact, we first invert a more general expression of the coordinate relations before applying them to the specific case of interest.  Thus, we consider:
\beq
\BO(\Bt,\Br)=|B|\left[\epsilon_{1}\cos(\Br)+\epsilon_{2}\cos(\Bt)\right],{\rm~with}\label{Og}
\eeq
\beq
R(r,t)={\sin(\Br)\over\BO},{\rm~and}\qquad T(\Bt,\Br)={\epsilon_{2}\sin(\Bt)\over\BO},\label{Cg}
\eeq
in which time orientation is preserved;  the quantities $\epsilon_{1},\epsilon_{2}$ may independently be $\pm1$, and the equations apply only in domains where $\BO\ge0$.  These can be inverted to give:
\beq
\tan(\Br)={\epsilon_{1}2|B|R\over B^2(T^2-R^2)+1},{\rm~and}\qquad \tan(\Bt)=-{2|B|T\over B^2(T^2-R^2)-1}.\label{cg}
\eeq
In the domain of part A, $\epsilon_{1}$ and $\epsilon_{2}$ are both $+1$.  It is interesting to note that the results in (\ref{cg}) are independent of $\epsilon_{2}$, so they apply unchanged for both signs indicated in (\ref{14}).  %Thus the expression for the coordinate inversion is identical for case A, and for the 
%It is straight forward
%to observe that $r=0\Leftrightarrow R=0$ for all $T$ or $t$. Calculations
%yield \begin{equation}
%r(T,R)=\frac{\sqrt{\left[\beta^{2}(T^{2}-R^{2})+1\right]^{2}+4\beta^{2}R^{2}}-\left[\beta^{2}(T^{2}-R^{2})+1\right]}{2R},\label{s1}\end{equation}
%\begin{equation}
%\tan^{2}(\beta t)=\frac{\beta^{2}R^{2}-Rr}{Rr(1+R\Br)}.\label{s2}\end{equation}
%Inserting (\ref{s1}) into (\ref{s2}) gives the functional form of
%$t(T,R)$, which is not very illuminating. Taylor expansion about
%$R=0$ gives \[
%r=\frac{\beta^{2}}{1+\beta^{2}T^{2}}R+\frac{\beta^{6}T^{2}}{\left(1+\beta^{2}T^{2}\right)^{3}}R^{3}+O(R^{5}),\]
%\[
%\tan^{2}(\beta t)=\beta^{2}T^{2}\left[1-\frac{2\beta^{2}}{\left(1+\beta^{2}T^{2}\right)^{2}}R^{2}+O(R^{4})\right].\]

\section*{A Point to remember}\label{unlabelled}

%It should be made clear that t
The conformal transformations %factor and the conformal coordinates 
in (\ref{20},\ref{23},\ref{25},\ref{26}) are not expressed
in terms of the original Schwarzschild coordinates $(\Bt,\, \Br)$ as
given in (\ref{1});  rather they are expressed in terms of the coordinates
$(\Bt,\, \Br)$ as given in (\ref{3-sphere}). Tracing back the coordinate transformations
performed in section \ref{section 1} and section \ref{section 2} gives the relation between the coordinates in (\ref{1}) and (\ref{3-sphere}). To obtain the conformal factor and the conformal coordinates in terms of the original Schwarzschild coordinates
in (\ref{1}), we have to multiply $\BO$ by $\To$ (see (\ref{3})) and by $\Bo$ (see (\ref{3-sphere})),
and replace $\Br$ and $\Bt$ by the following functions:
\begin{equation}
 \Bt%_{\ds}
=\sqrt{\alpha(a^2-1/4)}t%_{\Ss}
,{\rm~and}\qquad \Br%_{\ds}
=2\arctan\left({r%_{\Ss}
\over 1+\sqrt{1-\alpha r%_{Ss}
^2}}\sqrt{{\alpha(2a+1)\over(2a-1)}}\right).
%r\rightarrow\frac{2r\sqrt{\alpha}}{1+\sqrt{1-\alpha r^{2}}}\,\,;\,\, t\rightarrow\frac{2a+1}{8}t\sqrt{\alpha}.
\label{P1}
\end{equation}
%where subscripts refer to the {\bf 3}-{\bf s}phere and {\bf S}chwarzschild {\bf s}tar, respectively.

\section{Stellar Application:  electrostatics}\label{section 5}

Since Maxwell's equations are conformally invariant, and since the
Schwarzschild star metric in (\ref{schstr}) is conformally flat, any electromagnetic
problem inside the Schwarzschild star can be translated into an electromagnetic
problem in Minkowski space. We use the coordinate system $(\Bt,\Br,\theta,\phi)$
and the Schwarzschild star metric as in (\ref{3})  and (\ref{3-sphere}), and for the flat metric,
we shall use the coordinate system $(T,R,\theta,\phi)$ as given
in section \ref{section 4A}. We consider a static point charge $e$, inside
the Schwarzschild star at $\Br=\Br_{0}$. When we translate the problem
into flat spacetime electrodynamics, the charge will not be %considered
static. We describe its motion by the function $R=\Re(T,\Br_0)$, using 
%. To find this function, we use 
(\ref{cg}), without reference to $\Bt$: 
%to substitute for $\tan(t)$
%in (\ref{23}) to arrive at an expression for $T(R,r)$. Then fixing
%$r=r_{0}$ and inverting the function $T(R,r_{0})$ yields $\Re(T)$.
%\[
%T(R,r)=\frac{\sqrt{\left(\beta^{2}-r^{2}\right)R+\left(\beta^{2}R^{2}-1\right)r}}{\sqrt{r}\beta}\]
\begin{equation}
%\Rightarrow\Re(T)
\Re=\frac{-\cos(\Br_0)+\sqrt{1+\sin^2(\Br_0)T^{2}}}{\sin(\Br_0)}.\label{a1}
\end{equation}

The four velocity of the charge is given by 
\[
U^{\mu}=\frac{dT}{d\tau}\left(1,\frac{d\Re}{dT},0,0\right),\label{a0}
\]
where $\tau$ is the proper time of the charge. From (\ref{a1}),
we have
\begin{equation}
\frac{d\Re}{dT}=\frac{\sin(\Br_0)T}{\sqrt{1+\sin^2(\Br_0)T^{2}}}.\label{a2}
\end{equation}
Hence, $U^{\mu}$ can be expressed as
\begin{equation}
U^{\mu}=\left(\sqrt{1+\sin^2(\Br_0)T^{2}},\sin(\Br_0)T,0,0\right),\label{a3}
\end{equation}
in which the normalization has fixed $U^{\mu}U_{\mu}=-1$.
%which gives $c=\beta^{2}+r_{0}^{2}$. 
The proper time of the charge
$\tau$, can be obtained directly from (\ref{a3}): 
\begin{equation}
 \frac{dT}{d\tau}=\sqrt{1+\sin^2(\Br_0)T^{2}}\quad\Rightarrow\quad\tau=\frac{1}{\sin(\Br_0)}\ln\left[\sin(\Br_0)T+\sqrt{1+\sin^{2}(\Br_0)T^{2}}\right].\label{a4}
\end{equation}
%where $s\equiv\sin(\Br_0)$. In terms of
%$s$, the four velocity takes a simple form $U^{\mu}=\left(\sqrt{1+s^{2}T^{2}},sT,0,0\right)$.

The four acceleration $a^{\mu}$ of the charge is given by:
\begin{equation}
a^{\mu}=\frac{dU^{\mu}}{d\tau}=\frac{dU^{\mu}}{dT}\frac{dT}{d\tau}=\left(\sin^{2}(\Br_0)T,\, \sin(\Br_0)\sqrt{1+\sin^{2}(\Br_0)T^{2}},0,0\right).\label{a5}
\end{equation}
An interesting property of this acceleration is that its magnitude
is a constant, $a^{\mu}a_{\mu}=\sin^{2}(\Br_0)$. We end by concluding that
%indicating that this section by stating that, solving 
the problem of electrostatics inside the Schwarzschild star corresponds to
%geometry is tantamount to solving 
electrodynamics in a flat geometry
with the current density $J^{\mu}(T)=e\delta(R-\Re(T))\, U^{\mu}(T)/\sqrt{1+\sin^{2}(\Br_0)T^{2}}$.

\section{Robertson-Walker metric}\label{section 6}

%Let us now take a diversion toward 
We finish with a brief discussion of
the Robertson-Walker metric, which
is also a conformally flat metric (Weyl tensor evaluates to zero).  It often appears as a matter solution of the Einstein equations in contexts in which the Schwarzschild interior solution is also discussed, and it has similar conformal properties.  We have:
\begin{equation}
ds^{2}_{\RW}=\eta^{2}(t)\left[-dt^{2}+\frac{dr^{2}}{1-kr^{2}}+r^{2}d\Omega^2
%\left[d\theta^{2}+\sin^{2}\theta\, d\phi^{2}\right]
\right].\label{R1}
\end{equation}
Here $\eta(t)$ is the expansion factor of the universe and $k=0,\pm1$.
For $k=+1$ (spatially closed universe), the coordinate $r$ ranges
from $0$ to $1$. For $k=0$ or $-1$ (spatially open), the coordinate $r$
ranges from $0$ to $\infty$.

Following the procedure of sections \ref{section 2} through \ref{section 4}, we can find the conformal
coordinates of this metric. For $k=0$, these coordinates are by themselves
conformal coordinates. For $k=\pm1$, we can perform coordinate transformations
as in section \ref{section 1}, 
\begin{equation}
\Tlr=\frac{r}{1+\sqrt{1-kr^{2}}},{\rm~and}\qquad %\,,\, t/4\rightarrow 
\Tt=t/2,\label{R2}
\end{equation}
to obtain the metric in the form:
\begin{equation}
ds^{2}_{\RW}=\frac{4\eta^{2}(t)}{(1+k\Tlr^{2})^{2}}\left[-(1+k\Tlr^{2})^{2}d\Tt^{2}+d\Tlr^{2}+\Tlr^{2}d\Omega^2
%\left[d\theta^{2}+\sin^{2}\theta\, d\phi^{2}\right]
\right].\label{32}
\end{equation}
This is equivalent to (\ref{3}) with %the identification that $ds^{2}=\To^{2}d\Ts^{2}$, where 
$\To\equiv2\eta(t)/(1+k\Tlr^{2})$, %and 
$\beta=1$ and $0\le\Tlr\le1$, for $k=\pm1$.
%. For both values of $k$, the new coordinate $\Tlr$ ranges from $0$
%to $2$. %For convenience, we shall relabel $\Tlr$ as $r$.
Now, following the steps of sections \ref{section 2} and \ref{section 3}, we can obtain the conformal factor $\TO(\Tt,\Tlr)$
which makes the metric $\TO^{-2}d\Ts^{2}$ flat.

For $k\!=\!1$, we obtain, as previously, %exactly the same 
$\TO\!=\!\Bo\BO$, with $\Bo$ as in (\ref{3-sphere}) and $\BO$ as in (\ref{14}),
hence the same form for the conformal coordinates as in sections \ref{section 2} and \ref{section 4}. %So
The corresponding results for the Schwarzschild star thus apply exactly for
the $k=1$ case of the RW metric.

For $k\!=\!-1$, note that $g_{\Tt\Tt}\!=\!-(\beta^{2}-\Tlr^{2})$, which changes
the results from step (\ref{3-sphere}).  %In particular, w
We will distinguish between the two families of $\BO(\Bt,\Br)$ which ensue, %result, 
because of their rather different character relative to the result in (\ref{14}).  
They are related by a $\pm i\pi$ shift in $t_0$: %$t_0\rightarrow t_0\pm i\pi$:
%Here we obtain two distinct families
%of $\Omega(t,r)$ which cannot be together expressed as a single formula
%in terms of two arbitrary real constants $A,\, t_{0}$, as we did
%in (\ref{18})-\ref{19}. If we relax the requirement of the arbitrary
%constants being real, then the two families can be expressed by a
%single general formula, but since we need $\Omega(t,r)$ to be real
%(and positive), we stick to only real constants $A,\, t_{0}$. 
\begin{equation}
\BO_{1}(\Bt,\Br)=B\left[\cosh(\Br)+\cosh(\Bt-t_{0})\right],\label{R3}
\end{equation}
\begin{equation}
\BO_{2}(\Bt,\Br)=B\left[-\cosh(\Br)+\cosh(\Bt-t_{0})\right].\label{R4}
\end{equation}
%These are related %to each other 
%by a %$t_0\rightarrow t_0=
%$\pm i\pi$ in $t_0$.  
The time dependence is again of the form $t-t_{0}$ due to %reflecting %which again signifies
the time translation invariance of the RW metric. With %Choosing 
$B\!=\!1$ and
$t_{0}\!=\!0$ %in the first family of conformal factors $\Omega_{1}$ 
for $\BO_1$ given in (\ref{R3}),
and by following steps as in appendix \ref{missed}, %section \ref{section 4}, 
we obtain %end up with 
the following conformal coordinates:
\beq
 R(\Bt,\Br)=\frac{\sinh(\Br)}{\left[\cosh(\Br)+\cosh(\Bt)\right]},{\rm~and}\qquad
%\eeq
%\begin{equation}
T(\Bt,\Br)=\frac{\sinh(\Bt)}{\left[\cosh(\Br)+\cosh(\Bt)\right]}.\label{R5}
\end{equation}
Unlike the corresponding conformal coordinates for Schwarzschild star (section \ref{section 4}),
these conformal coordinates cover the entire spacetime, since this $\BO$ never vanishes. 
%This is because $\beta=2$, $r<2$ and $\left|\tanh()\right|<1$.

%On the other hand, 
By contrast, members of the %second 
family of conformal factors shown %$\Omega_{2}$ 
in (\ref{R4})
are not valid everywhere in the spacetime, and %. Nevertheless, 
as in the case of Schwarzschild star, this family $(\BO_{2})$, splits into
two subfamilies, one with $B>0$ and the other with $B<0$.  %, such that
For each $t_0$, their domains of validity are complementary, in that %to each other. That is, a pair (one from each subfamily with same $t_{0}$) would 
together, they 
cover the entire spacetime. Choosing $B\!=\!1$ and $t_{0}\!=\!0$ for $\BO_{2}$,
we obtain the following conformal coordinates:
\beq
 R(\Bt,\Br)=\frac{\sinh(\Br)}{\left[-\cosh(\Br)+\cosh(\Bt)\right]},{\rm~and}\qquad
%\eeq
%\begin{equation}
T(\Bt,\Br)=\frac{\pm\sinh(\Bt)}{\left[-\cosh(\Br)+\cosh(\Bt)\right]}.
%.\[R(t,r)=\frac{r}{\left[\beta^{2}-(\beta^{2}-r^{2})\cosh^{2}(\beta t)\right]},\]
%\begin{equation}
%T(t,r)=\frac{(\beta^{2}-r^{2})\coth(\beta t)}{\beta\left[\beta^{2}-r^{2}\coth^{2}(\beta t)\right]}.
\label{R6}
\end{equation}
Note: whereas (\ref{25}) and (\ref{26}) apply in the domain $|\Bt|<\Br$, (\ref{R6}) applies in $\Br<|\Bt|$.

\section{Discussion and Summary}\label{section 7}

We have constructed conformal coordinate systems for two conformally
flat geometries, the Schwarzschild interior solution and the Robertson-Walker
metric. Their conformal flatness is %becomes 
manifest in these coordinates.
%It is clearly shown that, 
Without involving the spherically symmetric 2-spheres, we show that these coordinate systems have two degrees
of freedom, one corresponding to a scaling factor $B$, and the other
corresponding to time translation $t_{0}$.

The Schwarzschild star metric (\ref{3}) and the RW metric (\ref{32})
for a closed universe ($k\!=\!1$) require identical treatment for 
%when it comes to 
finding %out 
their conformal coordinates. By explicitly %specifically
choosing $t_{0}$ and $B$, we arrive at specific choices for these
conformal coordinates, (\ref{20}, \ref{25}). These coordinates
cover only a part of the entire spacetime (see, for example (\ref{24})). %The
Complementary regions of the spacetime (%which are 
not covered by the chosen
coordinates) are covered by other coordinate systems %which
having the same $t_{0}$ as (\ref{20}, \ref{25}), but a negative
$B$. We consider these %two 
complementary coordinate systems 
%are hence considered
to belong to %two 
distinct subfamilies.

For the open universe ($k\!=\!-1$) RW metric, we obtain two families
of coordinate systems. A specific member from the first family is
given by (\ref{R5}). This coordinate system is %very 
well behaved
and covers the entire spacetime. %But, the m
Members of the second
family 
%of coordinate systems 
do not individually cover the entire
spacetime.  As in the $k=1$ case, 
they further split into two subfamilies, which are %such that 
%every coordinate system from one of these subfamilies will have a 
complementary in that
%coordinate system from the other subfamily, and %they would 
together they
cover the entire spacetime. A specific member from the second family
is given by (\ref{R6}).

A %very 
useful application of these conformal coordinates is in electrodynamics,
because %which is a theory guided by 
Maxwell's equations %which 
are conformally
invariant. In solving these equations, %we can get away with using just the 
it suffices to use flat space Green's functions, which simplify calculations
somewhat.  %considerably.  %a lot. Of course, w
We do have to expend some of the effort saved
in order to find the appropriate current density $J^{\mu}$, which requires further 
%we will have to do a suitable %(and more general) 
analysis.  In section \ref{section 5}, %(where we looked only at 
we examine the orbit of a static point source, that becomes dynamical in
%which depends on 
the flat %conformal
coordinates. %In ongoing work on 
For electromagnetic self-force calculations
in conformally flat geometries, these conformal coordinates are proving useful.

%\vfill\eject

\appendix

\section{Transformation from the 3-sphere to flat space}\label{missed}

We start from the 4-metric with spatial 3-sphere slices, with the form obtained in (\ref{3-sphere}):
\begin{equation}
ds^{2}=-d{t}^2+d{r}^2+\sin^2({r}) d\Omega^2,\label{A3-sphere}
\end{equation}
%Then, we 
and introduce %new, 
flat coordinates $T(t,r)$ and $R(t,r)$ as in (\ref{conf:eqn}).  
Expressing the differentials of the new coordinates in terms of the
old coordinates, we can write (\ref{A3-sphere}) as:
\begin{equation}
 \begin{array}{c}
ds^2=-%[\beta^{2}+r^{2}]^{2}
dt^{2}+dr^{2}+\sin^{2}(r)d\Omega^2
%\left(d\theta^{2}+\sin^{2}\theta d\phi^{2}\right)
\,=\\
\begin{array}{c}
%\,\,\,\,\,\,
\Omega^{2}(t,r)\left[-(\dot{T}^{2}-\dot{R}^{2})dt^{2}+(R'^{2}-T'^{2})dr^{2}%\right.\\
%\,\,\,\,\,\,\,\,\,\,\,\,\,\,\,\,\,\,\,\,\left.
+2(R'\dot{R}-T'\dot{T})dtdr+R^{2}d\Omega^2
%\left(d\theta^{2}+\sin^{2}\theta d\phi^{2}\right)
\right],\nonumber
\end{array}
\end{array}\label{5}
\end{equation}
where a dot denotes the derivative with respect to $t$ and a prime %dash
denotes the derivative with respect to $r$. Equating the coefficients
of identical differentials on either sides gives a set of relations in which $T$ does not appear undifferentiated:  %explicitly:
\begin{equation}
\begin{array}{cc}
1=\Omega^{2}(\dot{T}^{2}-\dot{R}^{2}), & I\\
\Omega^{2}(R'^{2}-T'^{2})=1, & II\\
\sin^{2}(r)=\Omega^{2}R^{2}, & III\\
\dot{T}T'=\dot{R}R',\hbox{\rm and thus} & IV\\
\dot{T}^2+T'^{2}=\dot{R}^2+R'^{2}, & V
\end{array}\label{6}
\end{equation}
where the last result follows after eliminating $\Omega$ from equations I and II.  Note:  these last two equations can be combined to give, by elementary algebra:
\begin{equation}
\begin{array}{ccc}
\dot{T}&=&\mp R',\\ %{\rm and}\\ %\qquad 
T'&=&\mp\dot{R},\label{missed1}
\end{array}
\end{equation}
where only results compatible with I and II have been retained.  
Further differentiation and elimination gives the 2-D wave equation for both $T$ and $R$, which because of the relations (\ref{missed1}), yeilds the solutions:
\begin{equation}
\begin{array}{ccc}
R(t,r)&=&f(t-r)+g(t+r),\hbox{\rm and}\\ %\qquad 
T(t,r)&=&\pm\left[f(t-r)-g(t+r)\right].\label{missed2}
\end{array}
\end{equation}
Elimination of $\Omega$ from I and III, and substitution of (\ref{missed2}) into the result, gives:
\begin{equation}
(f+g)^2+4\dot{f}\dot{g}\sin^{2}(r)=0,\label{missed3}
\end{equation}
while equations I and II together directly give:
\begin{equation}
-4\dot{f}\,\dot{g}=\Omega^{-2}=4f'g',\label{missed4}
\end{equation}
indicating that $f'$ and $g'$ must have everywhere the same sign.

By taking the $t$ derivative of (\ref{missed3}) and eliminating $\sin^2(r)$ from the result, we obtain an equation which can be rearranged to give:
\begin{equation}
\left(2\dot{f}-{f\ddot{f}\over\dot{f}} \right)-\left(g{\ddot{f}\over\dot{f}}+f{\ddot{g}\over\dot{g}}\right)+\left(2\dot{g}-{g\ddot{g}\over\dot{g}} \right)=0.\label{missed5}
\end{equation}
Futher differentiation, once by the argument of $f$ and once by the argument of $g$, then rearranging to separate variables, gives (with $\alpha$ a constant of separation):  
\begin{equation}
{1\over\dot{f}}\left({\ddot{f}\over\dot{f}} \right)^{.}=-{1\over\dot{g}}\left({\ddot{g}\over\dot{g}} \right)^{.}=-\alpha,\label{missed6}
\end{equation}
which can be solved %integrated 
completely.  Integrating once gives (with integration constants $a, c$):
\begin{equation}
{\ddot{f}\over\dot{f}}=-\alpha(f+a), {\rm~and}\qquad{\ddot{g}\over\dot{g}}=\alpha(g+c).\label{missed7}
\end{equation}
Substitution of these back into (\ref{missed5}) allows further separation of variables (with separation constant $k$):
\begin{equation}
2\dot{f}-{f\ddot{f}\over\dot{f}}-\alpha cf=-2\dot{g}+{g\ddot{g}\over\dot{g}}-\alpha ag=-k,\label{missed8}
\end{equation}
while further integration of (\ref{missed7}) gives (with new integration constants $b,d$):
\begin{equation}
\dot{f}=-{\alpha\over 2}\left[(f+a)^2+b^2\right],{\rm~and}\qquad\dot{g}={\alpha\over 2}\left[(g+c)^2+d^2\right].\label{missed9}
\end{equation}
Substitution of (\ref{missed9}) and (\ref{missed7}) back into (\ref{missed8}) allows us to conclude:
\begin{equation}
c=-a,\qquad d=\pm b,{\rm~and}\qquad k=\alpha(a^2+b^2).\label{missed10}
\end{equation}
Finally, further integration of (\ref{missed9}) yields (with additional integration constants $e,h$):
\begin{equation}
 f=-a+b\cot{{\alpha b\over 2}(t-r+e)},{\rm~and}\qquad g=a-b\cot{{\alpha b\over 2}(t+r+h)},\label{missed11}
\end{equation}
unless $b=0$ (relevant for the flat space treatment in appendix \ref{appA}), in which case:
\begin{equation}
f=-a+{2\over \alpha(t-r+e)},{\rm~and}\qquad g=a-{2\over \alpha(t+r+h)}.\label{missed12}
\end{equation}
Up to this point nothing has depended on the metric coefficient multiplying the metric on the 2-sphere.  Now, using the solution (\ref{missed11}) back in (\ref{missed3}) where that coefficient appears, and simplifying, gives us:
\begin{equation}
\alpha b\sin(r)=\pm\sin{\alpha b\left(r+{h-e\over 2}\right)},\label{missed13}
\end{equation}
from which we conclude (representing time translation invariance through the offset $t_0$):
\begin{equation}
\alpha b=1,\qquad e=-t_0-n\pi,{\rm~and}\qquad h=-t_0+n\pi,\label{missed14}
\end{equation}
where $n$ is an integer ($+$ve or $-$ve), even for the $+$ sign in (\ref{missed13}), odd for the $-$ sign.  We can now solve for $R,~T$ (after replacing $a$ by $\mp T_0/2$) and $\Omega$:  
\begin{eqnarray}
R(t,r)&=&{2\over\alpha}{\sin(r)\over\cos(r)-(-1)^{n}\cos(t-t_0)},\nonumber
\\
T(t,r)&=&{2\over\alpha}{\pm\sin(t-t_0)\over\cos(r)-(-1)^{n}\cos(t-t_0)}+T_0,\label{missed15}\\
\Omega(t,r)&=&{\alpha\over 2}\left[\cos(r)-(-1)^{n}\cos(t-t_0)\right].
\nonumber
\end{eqnarray}
Note:  we must choose $\alpha$ so that $R,\Omega>0$, we may, in principle,  absorb the $(-1)^{n}$ into a redefinition of $t_{0}$, and we would want %to choose 
the $\pm$ sign %so as 
to preserve time orientation.

\section{Transformations within flat space}\label{appA}

We consider the flat space metric in spherical polar coordinates and find the most general, spherically symmetric, conformal transformation
$H$, that maintains the flatness of this metric. In other words,
we find $H(T,R)$, and the coordinate functions $R'(T,R)$ and $T'(T,R)$, 
which satisfy the following equation: 
\begin{equation}
\left[-dT^{2}+dR^{2}+R^{2}d\Omega^{2}\right]=H^{2}\left[-dT'^{2}+dR'^{2}+R'^{2}d\Omega^{2}\right].\label{A1}
\end{equation}
By equating the components of the Riemann tensor of the above metric
to zero, we obtain equations similar to those in section3 \{eq(i)-(v)\},
with $\sin(r)\rightarrow r$ and $\Omega$ replaced by $H$. These equations can
be solved in a way similar to that used %what has been done 
in section \ref{section 3}, and the most general solution is found to be: 
\begin{equation}
\begin{array}{c}
H(T,R)=D\left[(T-T_{0})^{2}-R^{2}\right],\\
T'-T'_{0}=\mp(T-T_{0})/H,{\rm~and}\\
R'=R/H,
\end{array}
\label{A2}
\end{equation}
where $D$ and $T_{0}$ are arbitrary parameters.  
Null infinity becomes the lightcone emerging from $T'_0$.  
The new coordinates 
%system 
$\{R',T'\}$ and the conformal factor $H$ are singular %blows up 
on the light cone emerging from the point %$\{R,T\}=\{0,T_0\}$.  
$R=0,T=T_{0}$.  
We refer to 
%Let us call 
this light cone as $L(T_{0})$.  

\begin{figure}%[htbp]  
\scalebox{0.60}{ \psfig{file=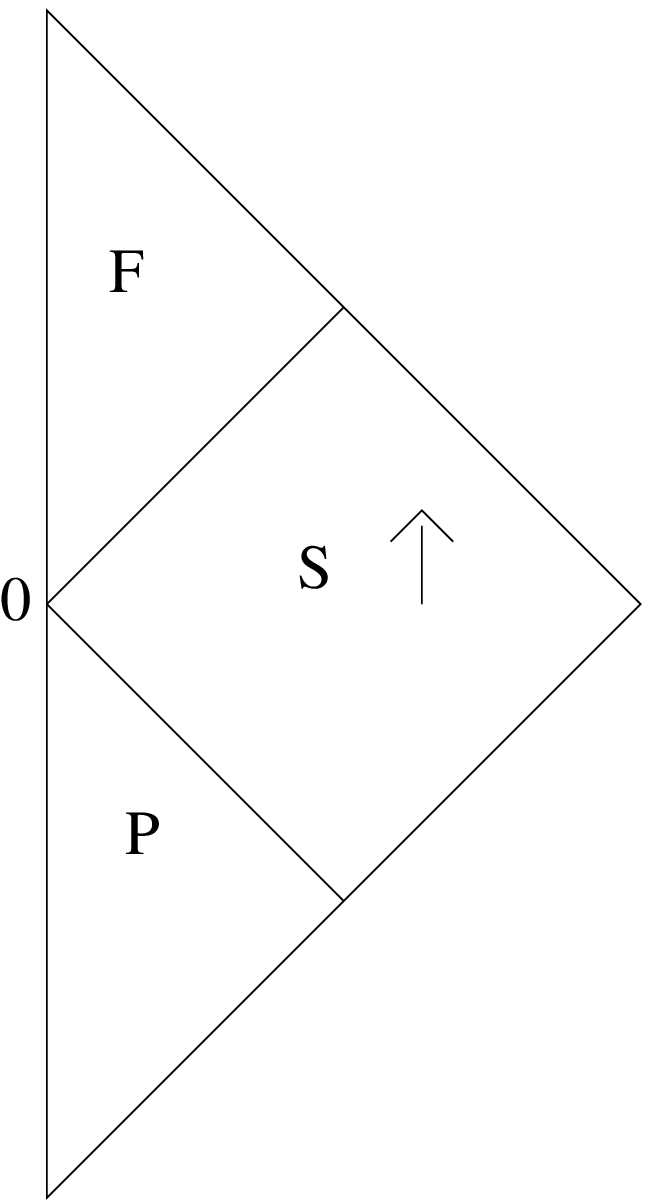} }
~~~~~~~~\,~~~~~~~~
\scalebox{0.67}{\psfig{file=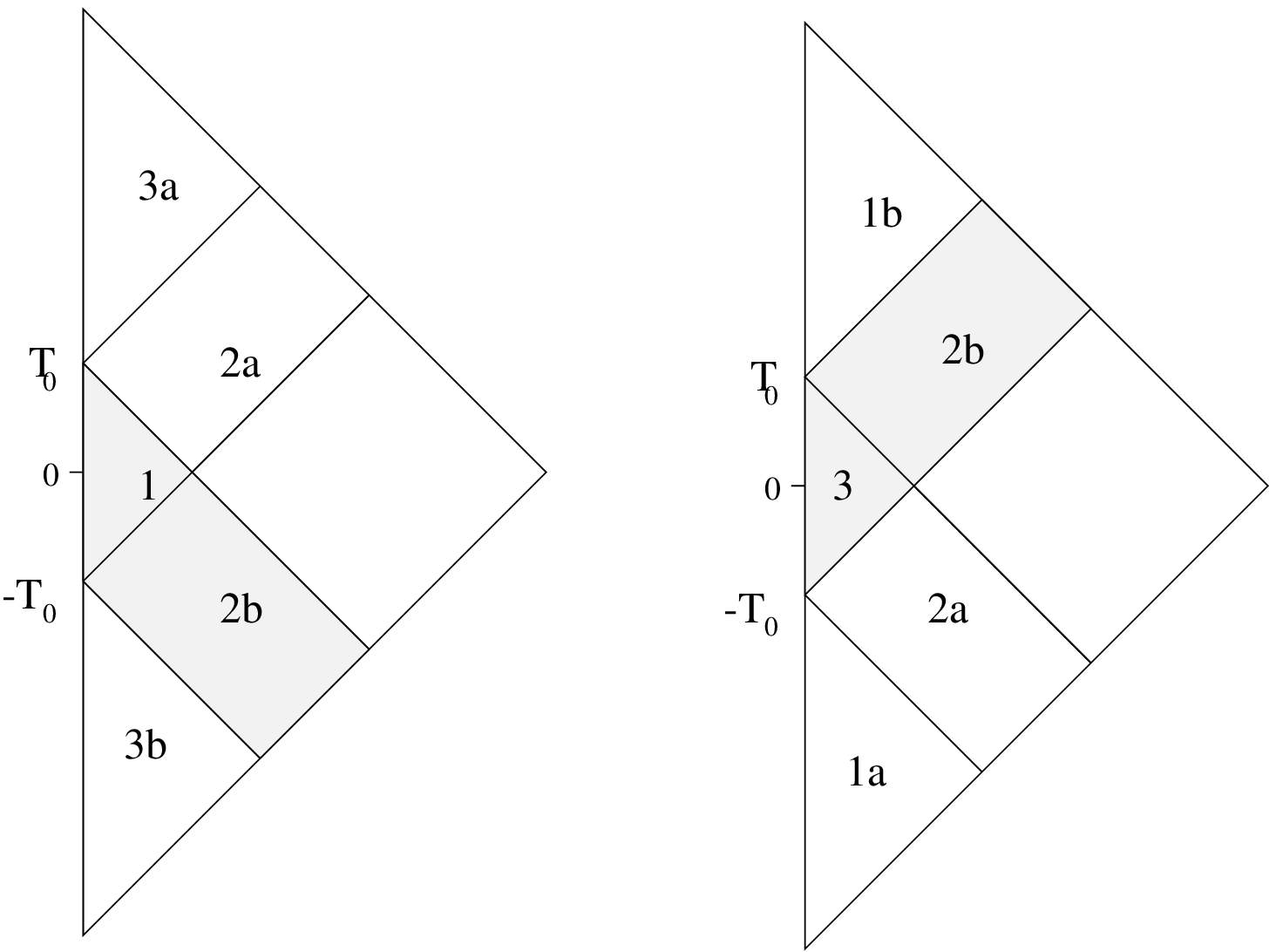} }
\caption{At the left, Minkowski space is shown, with Future, Past and Spacelike regions (relative to the origin) labeled, and time-orientation indicated.  In the middle, the light cones above/below $\mp T_0$ map to the light cones below/above $\pm T_0$ on the right.  Regions 2a and 2b in the middle map to regions with the same name on the right, and neighboring regions map accordingly.  The type of map which takes 1a on the right directly to 1b on the right is characterized by (\ref{A4}).}
\label{fig1}  
\end{figure} 

The requirement that the conformal factor be positive
definite $(H>0)$, splits the family of conformal factors (parameterised
by $D$ and $T_{0}$) naturally into two subfamilies, one with
$D>0$ and the other with $D<0$. The absolute magnitude of $D$ corresponds
to a relatively trivial scaling factor which can be absorbed into the coordinate
functions. When $D<0$, the conformal factor in (\ref{A2}) is valid only in the
elsewhere region of the light cone $L(T_{0})$, while with $D>0$, it
is valid only within the causal regions of $L(T_{0})$ (see left panel of Fig \ref{fig1}). These
two subfamilies are complementary %to each other, because they both
in that together they span the entire space time.  Note that, under this %coordinate
transformation, all the points on
the light cone $L(T_{0})$ map to infinity --- more specifically, to the boundary of the conformal completion of Minkowski space. 
Hence, choosing a specific coordinate transformation first
involves choosing a specific light cone $L(T_{0})$, and then %secondly
choosing to span either its causal region or its elsewhere region.
%Observe that, t
The $\pm$ sign in front of the time coordinate corresponds
to the time reversal symmetry intrinsic to the metric. An appropriate
sign can be chosen by requiring that $T'$ is a monotonically increasing
function of $T$. Thus, when $D>0$, we shall choose the $-$ sign, and Future and Past map into each other.  When
$D<0$, we shall choose the $+$ sign, and map spacelike Elsewhere into itself.
 
We now consider two different light cones $L(T_{01})$ and $L(T_{02})$.
Correspondingly, we choose two transformations $\{H_{1},T',R'\}$
and $\{H_{2},\widetilde{T},\widetilde{R}\}$. Clearly, in the region of overlap
of these two transformations, the new coordinate functions
can be expressed in terms of one another. That is, we can express
$\{H_{2},\widetilde{T},\widetilde{R}\}$ as a function of $\{T',R'\}$. The generality
of (\ref{A2}) ensures that this new transformation is of
the same form as (\ref{A2}). We find the transformation explicitly
and show its conformance with (\ref{A2}).

%Without loss of generality, we can choose one of the light cones to emanate from $R=T=0$, and the other to emanate from $R=0,T=T_{0}$.  Also, let us 
We choose the two coordinate systems $(T',R')$ and $(\widetilde{T},\widetilde{R})$
to span the causal region within the respective light cones: 
\begin{equation}
 \begin{array}{cc}
\left\{ \begin{array}{c}
H_{1}=D_1\left[(T-T_{01})^{2}-R^{2}\right],\\
T'-T_{01}'=-(T-T_{01})/H_{1},\\
R'=R/H_{1},\end{array}\right\}  & {\rm and}\qquad \left\{ \begin{array}{c}
H_{2}=D_2\left[(T-T_{02})^{2}-R^{2}\right],\\
\widetilde{T}-\widetilde{T}_{02}=-(T-T_{02})/H_{2},\\ %+\widetilde{T}_{0},\\
\widetilde{R}=R/H_{2}.\end{array}\right\} \end{array}\label{A3}
\end{equation}
%As expected, it turns out that i
In the region of overlap of the two
coordinate systems (region 1 in the middle panel of Fig \ref{fig1}) $\widetilde{T}$ and $\widetilde{R}$ can be expressed
in terms of $T',R'$ (regions 1a and 1b in the panel on the right of Fig \ref{fig1}).  Note that time orientation has been preserved:
\begin{equation}
 \begin{array}{cc}
\left\{ \begin{array}{c}
H=\widetilde{D}\left[(T'-T_{0}')^{2}-R'^{2}\right],\\
\widetilde{T}-\widetilde{T_{0}}'=-(T'-T_{0}')/H,\\ %+(\widetilde{T}_{0}-1/T_{0})\\
\widetilde{R}=R'/H,
\end{array}\right\} & {\rm where}\quad
\left\{ \begin{array}{c}
\widetilde{D}=D_1D_2(T_{01}-T_{02})^2,\\
T_{0}'=T_{01}'+1/\left[D_1(T_{01}-T_{02})\right],\\
\widetilde{T}_{0}'=\widetilde{T}_{02}-1/\left[D_2(T_{01}-T_{02})\right].
\end{array}\right\}
\end{array}\label{A4}
\end{equation}
This is clearly seen to be of the same form as (\ref{A2}). The
properties of the conformal factor $H(T,R)$ demonstrated in this appendix
are used in section \ref{flatcomp} %the paper 
to elucidate the properties of the conformal
factors $\Omega(t,r)$ obtained in (\ref{14}) and (\ref{missed15}).

\end{document}